\pdfoutput=1
\documentclass[12pt]{article} 
\usepackage{geometry}
\usepackage{amsmath} 
\numberwithin{equation}{section}

\RequirePackage[pdftex,dvipsnames]{xcolor}
\definecolor{darkblue}{rgb}{0.1,0.1,.7}
\usepackage[colorlinks, linkcolor=darkblue, citecolor=darkblue, urlcolor=darkblue, linktocpage]{hyperref} 

\usepackage[square, comma, sort&compress,numbers]{natbib} 

\usepackage[margin=10pt,font=small,labelfont=bf]{caption}

\usepackage{tikz}
\usetikzlibrary{external}

\usepackage[]{sashorthand} 

\usepackage[final]{pdfpages}

\DeclareMathAlphabet{\mathpzc}{OT1}{pzc}{m}{it} 
\graphicspath{ {images/} }
\begin{document}
	\vspace*{-.6in} \thispagestyle{empty}
	\begin{flushright}
	\end{flushright}
	\vspace{.2in} {\Large
		\begin{center}
			{\bf Towards the higher point holographic momentum space amplitudes II: Gravitons\vspace{.1in}}
		\end{center}
	}
	\vspace{.2in}
	\begin{center}
		{\bf 
			Soner Albayrak$^{a,b}$ and Savan Kharel$^{a,c}$}
		\\
		\vspace{.2in} 
		$^a$ {\it  Department of Physics, Yale University, New Haven, CT 06511}\\
		$^b$ {\it  Walter Burke Institute for Theoretical Physics, Caltech, Pasadena, CA 91125}\\
		$^c$ {\it  Department of Physics, Williams College, Williamstown, MA 01267}
	\end{center}
	
	\vspace{.2in}
	
\begin{abstract}
In this follow up paper, we calculate higher point tree level graviton Witten diagrams in AdS$_4$ via bulk perturbation theory. We show that by rearranging the bulk to bulk graviton propagators, the calculations effectively reduce to the computation of a scalar factor. Analogous to the amplitudes for vector boson interactions we computed in the previous paper, scalar factors for the graviton exchange diagrams also become relatively simple when written in momentum space. We explicitly calculate higher point correlators and discuss how this momentum space formalism makes flat space and collinear limits simpler.
\end{abstract}
	
	\newpage
	
	\tableofcontents
	
	
\section{Introduction}
 The AdS/CFT correspondence maps gravitational theories with their non-gravitational counterparts
 \cite{Maldacena:1997re, Witten:1998qj}. This correspondence has provided a concrete tool to tackle problems in theoretical physics from the nature of black holes to non-equilibrium phenomena in strongly coupled systems including condensed matter physics.\footnote{For a more recent introduction to AdS/CFT, see \cite{Penedones:2016voo}. For a complementary review on conformal field theories, see \cite{Poland:2018epd}.} 

In Minkowski space, the S-matrix provides the transition amplitude for a set of particles in an asymptotic \emph{in} state at $t =-\infty$ that turns into a different set of \emph{out} states at \mbox{$t = +\infty$}.\footnote{The standard Haag-Ruelle construction of S-matrix \cite{Haag:1958vt,Haag:1959ozr,ruelle1962asymptotic} assumes a mass gap in the spectrum hence S-matrix is not rigorously defined for massless particles. For a recent short discussion with an improved construction, see \cite{Hannesdottir:2019rqq}.} In an AdS with a time-like boundary such notions of in and out states are not very sensible. Instead, one can view the AdS as a box where particles can interact perpetually.  However, one can also alter the boundary conditions at the time-like boundary thereby creating and annihilating particles. We know from AdS/CFT correspondence that the transition amplitudes between these type of states are equal to the correlation functions of the dual conformal field theory (CFT) \cite{Penedones:2010ue}. Such amplitudes are among the most fundamental objects, and many important observables of the theories living in AdS are constructed through them. Because of their importance, they have garnered appreciable interest in the last decade.\footnote{For instance, early work in this direction was pioneered in these papers: \cite{Freedman:1998bj, Liu:1998ty, Freedman:1998tz, DHoker:1999mqo,DHoker:1999kzh}.} 

The study of the scattering amplitudes of gauge theories and gravity has revealed remarkably simple structures in flat space and has led to menagerie of basis such as twistors and geometric formulations like the amplituhedron \cite{Britto:2005fq, Witten:2003nn, ArkaniHamed:2008gz,ArkaniHamed:2009dn, ArkaniHamed:2009vw}. Additionally, in the last several years we have witnessed inspiring representations which have helped in showing similarities between scattering amplitudes in flat space and their AdS counterparts \cite{ Penedones:2010ue, Paulos:2011ie,  Mack:2009gy, Fitzpatrick:2011ia,Kharel:2013mka, Fitzpatrick:2011hu, Fitzpatrick:2011dm, Costa:2014kfa, Jepsen:2018ajn,Jepsen:2018dqp, Gubser:2018cha, Hijano:2015zsa, Yuan:2018qva, Parikh:2019ygo, Jepsen:2019svc}. One related insight has come out of the recent investigation of holographic momentum space which indicates an interesting connection between momentum space Witten diagrams and flat space scattering amplitudes. Indeed, one can obtain the S-matrix from the AdS correlation function with an elegant limit $k_1+ k_2+ \cdots k_n \mapsto 0$ where $k_i$ is the magnitude of the external momenta \cite{Raju:2012zr,Raju:2012zs}. A similar flat space limit also exist in de Sitter space \cite{Maldacena:2011nz,Arkani-Hamed:2015bza}. In fact, anti-de Sitter and de Sitter correlators are closely related and this intimacy provides a cosmological motivation for studying AdS amplitudes as well.\footnote{Even though mathematical framework of AdS and dS are intuitively related, their connection can be obscure. For example, it is shown in a recent paper \cite{Albayrak:2019asr} that an algorithm for the conformally coupled scalars in dS can be directly used for gauge bosons in AdS, a relation which is not manifestly obvious.} In that context, we are interested in studying the late time spatial correlations that encode the primordial scattering processes. 

While many exciting directions have been explored in the study of conformal structure in momentum space \cite{Raju:2010by, Mata:2012bx, Raju:2012zs, Raju:2012zr, Raju:2011mp, Arkani-Hamed:2015bza, Bzowski:2013sza, Bzowski:2015pba,Bzowski:2018fql, Bzowski:2015yxv, Maldacena:2002vr,Maldacena:2011nz, Isono:2019ihz, Isono:2018rrb, Coriano:2018bbe, Coriano:2019dyc, Maglio:2019grh, Gillioz:2018mto, Arkani-Hamed:2018kmz, Coriano:2018bsy, Coriano:2013jba}, a lot of of work still remains to be done. In our previous paper, we used momentum space formalism of \cite{Raju:2011mp} to compute the explicit expressions for several higher point vector correlators which take surprisingly simple forms \cite{Albayrak:2018tam}. Furthermore, it is shown in \cite{Albayrak:2019asr} that these calculation can be considerably simplified in a judiciously chosen basis. Specifically, different coefficients of the terms in the tensor structure of a vector amplitude are all related to a specific one among them, which means that the whole calculation reduces to only one integration.\footnote{In \cite{Albayrak:2019asr}, the authors also introduce an algebraic algorithm which bypasses the integration, which is extended to scalars in \cite{Albayrak:2020isk}. Extension of this algorithm to gravitons is an open problem that we would like to return later; however, in this paper, we will stick to explicit integration.} In this follow-up paper, we will discuss an analogous method to reduce the momentum space graviton calculations to computation of a scalar factor, and we will explicitly compute tree level higher point Witten diagrams with exchanged gravitons. Furthermore, we will discuss the flat space and collinear limits in our settings and provide explicit results. We think that our results could serve as data points from which further insights can be drawn. For instance, the inflationary cosmology has stimulated a great deal of excitement in the study of late time de Sitter correlators \cite{Guth:1980zm,Linde:1981mu,Albrecht:1982wi,Starobinsky:1982ee} and we believe that the analogous calculations of momentum space AdS amplitudes can assist in the study of the shape of non-Gaussianities  \cite{McFadden:2009fg, Chen:2009zp, Maldacena:2011nz, Baumann:2011nk, Assassi:2012zq,Chen:2012ge,Assassi:2013gxa,Arkani-Hamed:2015bza,Lee:2016vti,An:2017hlx,Kehagias:2017cym,Kumar:2017ecc,Baumann:2017jvh,Franciolini:2017ktv,Arkani-Hamed:2018kmz,Goon:2018fyu,Anninos:2019nib,Pi:2012gf,Gong:2013sma}.

Here is the organization of the paper. In section \ref{sec:2}, we briefly summarize momentum space perturbation theory of scalars and vector bosons, and present an overview of the gravitons. We also discuss how to strip off the tensor part of the graviton bulk to bulk propagator and how to effectively compute the remaining scalar factor for any tree level Witten diagram by introducing \emph{bulk point integrated expressions}. In section \ref{sec:3}, we use these ingredients and explicitly compute three, four  and five point functions. In section \ref{sec: limits} we obtain the expected flat space expressions, and further comment on how specific collinear limits can be simplified in our construction. Finally, we comment on many promising directions in the conclusion.

\section{Preliminaries}
\label{sec:2}

\subsection{\mbox{Review: AdS Momentum Space Perturbation Theory}}
In this section we provide a brief review of momentum space bulk perturbation theory. We refer reader interested in scalars and vector bosons to \cite{Albayrak:2018tam} and the references therein; in this follow-up paper, we will present a very succinct overview with an emphasis on the gravitons.

In order to write a momentum space amplitude for a tree level Witten diagram, one needs to take the product of all relevant bulk to bulk and bulk to boundary propagators with the vertex factors, followed by an integration along the bulk radial direction.\footnote{We work with the coordinates $\{z,k_i\}$ where $k_i$ are the Fourier transform of $x_i$ for the Poincar\'e patch $ds^2=z^{-2}\left(dz^2+\eta_{ij}dx^idx^j\right)$. Therefore what we call \emph{momentum space amplitude} is the amplitude in the coordinates $\{k_i\}$, which is obtained after the dependence on the bulk radius direction $z$ is integrated. 
} In the case of gluons in the axial gauge, the relevant ingredients for AdS$_{d+1}$ are
\bea[eq: vector ingredients] 
A_i^a(\bm{k},z)={}& \bm{\epsilon}_i^a \sqrt{\frac{2}{\pi}}(k z)^{{\frac{d-2}{2}}} K_{\frac{d-2}{2}}(k z)\;,\\
\mathcal{G}_{ij}(\bm{k}; z, z') ={}& (zz')^{{\frac{d-2}{2}}}\int\limits_{0}^{\infty} d\omega  J_{\frac{d-2}{2}}(\omega z)\frac{\omega H_{ij}^{(\omega,\bm{k})}}{ (\bm{k}^2+\omega^2-i \epsilon)} J_{{\frac{d-2}{2}}}(\omega z')\;,\\
H_{ij}^{(\omega,\bm{k})}\coloneqq{}&-i\left(\eta_{ij}+\frac{\bm{k}_i\bm{k}_j}{\omega^2}\right)\;,\\
\label{eq: vector boson 3 point vertex factor}
V_{ijk}(\bm{k}_1, \bm{k}_2, \bm{k}_3)\coloneqq{}&{}\frac{i}{\sqrt{2}}\left(\eta_{ij}(\bm{k}_1-\bm{k}_2)_k+\eta_{jk}(\bm{k}_2-\bm{k}_3)_i+\eta_{ki}(\bm{k}_3-\bm{k}_1)_j\right)\;,\\
V^{ijkl}_c\coloneqq{}&{} i \;\eta^{ik} \eta^{jl}-\frac{i}{2}\left(\eta^{ij} \eta^{kl}+\eta^{il} \eta^{jk}\right)\;.
\eea 
for the bulk to boundary propagator $A_i^a$, the bulk to bulk propagator $\mathcal{G}_{ij}$,\footnote{The integration range in the bulk-to-bulk propagator follows from a Bessel function identity used in its derivation, i.e. $\int_0^{\infty}J_\nu(a t)J_\nu(b t)tdt=a^{-1}\delta(a-b)$.} and the vertex factors $V$'s. In the equations above, $K_\nu(x)$ is the modified Bessel function of the second kind, $\eta_{ij}=\eta^{ij}$ is the boundary metric, $\bm{\epsilon}_i^a$ is the transverse polarization tensor ($\bm{\epsilon}_i^a\bm{k}^i=0$), and $k$ is the norm of the external momenta, i.e. $k=\sqrt{\abs{\bm{k}^2}}$. In this paper, we will stick to spacelike momenta, i.e. $\bm{k}^2=k^2$, as was done in \cite{Albayrak:2018tam}: one can analytically continue to timelike momenta, however the bulk to boundary propagators need to be modified accordingly \cite{Raju:2011mp}.

As an example, the color-ordered expression for the $s-$channel four point Witten diagram is written as
\begin{multline}
\label{eq: gluon 4pt}
\cM_{4s}=\int\limits_{0}^{\infty}\frac{dz}{z^{d+1}}\frac{dz'}{z'^{d+1}}A_i(\bm{k}_1,z)A_j(\bm{k}_2,z)V^{ijk}(\bm{k}_1, \bm{k}_2,-\bm{k}_1-\bm{k}_2)\mathcal{G}_{kl}(\bm{k}_1+\bm{k}_2; z, z')\\
\x z^4z'^4 V^{lmn}(\bm{k}_1, \bm{k}_2,\bm{k}_1+\bm{k}_2)A_m(\bm{k}_3,z')A_n(\bm{k}_4,z') \;.
\end{multline}
An explicit expression for $d=3$ for the above expression can be found in \cite{Albayrak:2018tam}.

In the expression above, we included the factor $z^4z'^4$. This was given as part of the overall prescription in \cite{Albayrak:2018tam} without a detailed explanation. It basically follows from the fact that the relevant inverse metric is $g^{ij}=z^2\eta^{ij}$ and both in three point and contact vertices we need two inverse metrics. So, to ease the notation in the vector calculations, we effectively took $g_{ij}=g^{ij}=\eta_{ij}=\eta^{ij}$ and inserted necessary $z$ factors at the end.\footnote{For a similar discussion, see section 6.2.2 of \cite{Raju:2011mp}.} In this paper, we present the tree point vertex factor and the propagators in contravariant and covariant forms respectively, so one does not need to worry about any additional $z-$factors.

The analogous set of ingredients to \equref{eq: vector ingredients} for gravitons in axial gauge is given by \cite{Raju:2011mp}:
\bea[eq: graviton ingredients] 
h_{ij}(\bm{k},z)={}& \epsilon_{ij}\sqrt{\frac{2}{\pi}}z^{-2}(k z)^{\frac{d}{2}} K_{\frac{d}{2}}(k z)\;,\\
\label{eq: graviton propagator}
\mathcal{G}_{ab,cd}(\bm{k}; z, z')={}&
\frac{i(zz')^{\frac{d}{2}-2}}{2}\int\limits_{0}^{\infty} d\omega 
J_{\frac{d}{2}}(\omega z)J_{\frac{d}{2}}(\omega z')\nn\\&{}\x\frac{\omega \left(H^{(\omega,\bm{k})}_{ac}H^{(\omega,\bm{k})}_{bd}+H^{(\omega,\bm{k})}_{ad}H^{(\omega,\bm{k})}_{bc}-\frac{2}{d-1}H^{(\omega,\bm{k})}_{ab}H^{(\omega,\bm{k})}_{cd}\right)}{k^2+\omega^2-i \epsilon}
\;,\\
\cV^{ijklmn}_{\bm{k}_1, \bm{k}_2, \bm{k}_3}\coloneqq{}&{} z^8V^{ijklmn}_{\bm{k}_1, \bm{k}_2, \bm{k}_3}\;,\\
V^{ijklmn}_{\bm{k}_1, \bm{k}_2, \bm{k}_3}\coloneqq{}&{} \left(
\frac{(\bm{k}_2)^i(\bm{k}_3)^j\eta^{km}\eta^{ln}}{4}-\frac{(\bm{k}_2)^i(\bm{k}_3)^k\eta^{jm}\eta^{ln}}{2}
\right)+\text{ permutations}\;.
\eea 
where $\epsilon_{ij}$ is the symmetric traceless transverse polarization tensor, e.g. $\epsilon_{ij}\bm{k}^i=\epsilon_{ij}\eta^{ij}=0$ and $\epsilon_{ij}=\epsilon_{ji}$. We defined $V^{ijklmn}_{\bm{k}_1, \bm{k}_2, \bm{k}_3}$ for convenience while $\cV^{ijklmn}_{\bm{k}_1, \bm{k}_2, \bm{k}_3}$ is the appropriate three point vertex factor.

The permutations in the vertex are generated by the permutation group element\linebreak $(\bm{k}_1\bm{k}_2\bm{k}_3)(ikm)(jln)$ in cycle notation.\footnote{See section 3.2.1. of \cite{Raju:2012zs} for the full contracted expression.} This is analogous to the vertex factor for the vector boson in \equref{eq: vector boson 3 point vertex factor} where the second and third terms can be obtained from the first one by the permutation $(\bm{k}_1\bm{k}_2\bm{k}_3)(ijk)$.

\subsection{Stripping off the tensorial part of the propagator}
Our goal in this part is to provide a prescription to simplify the propagator.
The form of the graviton propagator as given in \equref{eq: graviton propagator} shows that there are three different radial integrations we need to consider due to $\omega$ dependence of $H^{(\omega,\bm{k})}_{ac}H^{(\omega,\bm{k})}_{bd}$; however, it is not immediately clear if these integrations are related. In \cite{Albayrak:2019asr}, the authors addressed a similar problem for the gluon propagator. By rewriting the gluon propagator as
\small
\be
\mathcal{G}_{ij}(\bm{k}; z, z')=
\left(\eta_{ij}-\frac{\bm{k}_i\bm{k}_j}{k^2}\right)
\int
\omega d\omega 
\frac{\sqrt{z z'} J_{\frac{1}{2}}(\omega z) J_{\frac{1}{2}}(\omega z')}{i(k^2+\omega^2-i \epsilon)}
+
\frac{\bm{k}_i\bm{k}_j}{k^2}
\int
d\omega
\frac{k^2+\omega^2}{\omega} 
\frac{\sqrt{z z'} J_{\frac{1}{2}}(\omega z) J_{{\frac{1}{2}}}(\omega z')}{i(k^2+\omega^2-i \epsilon)}\;,
\ee
\normalsize
they introduced a decomposition of it; diagrammatically,
\begin{equation}
\includegraphics[scale=1.1]{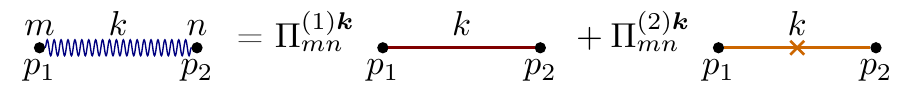}
\label{eq: propagator decomposition}
\end{equation}
for
\begin{equation}
\Pi^{(1)\bm{k}}_{mn}\equiv{}{}-i\left(\eta_{mn}-\frac{\bm{k}_m\bm{k}_n}{k^2}\right)\;,\quad
\Pi^{(2)\bm{k}}_{mn}\equiv{}{}-i \frac{\bm{k}_m\bm{k}_n}{k^2}\;,
\label{eq: projectors}
\end{equation}
and
\be 
\label{eq: relation of straight and crossed lines}
\begin{aligned}
	\includegraphics[scale=1.2]{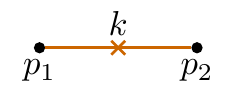}
\end{aligned}=\lim\limits_{k\rightarrow 0}\begin{aligned}
	\includegraphics[scale=1.2]{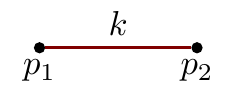}
\end{aligned}
\ee 
where
\be 
\begin{aligned}
	\includegraphics[scale=1.2]{straight}
\end{aligned}\coloneqq 
\int
\omega d\omega 
\frac{\sqrt{z z'} J_{\frac{1}{2}}(\omega z) J_{\frac{1}{2}}(\omega z')}{k^2+\omega^2-i \epsilon}\;.
\ee 
Using this decomposition, instead of performing $2^n$ integrations where $n$ is  number of bulk to bulk propagator, one can only proceed with one explicit integration and obtain the rest via \equref{eq: relation of straight and crossed lines}.

We are actually abusing the notation in the equations above as these lines stand for \emph{bulk-point integrated} diagrams in \cite{Albayrak:2019asr}. In the case of gluons, the additive property of the norm of momenta of the bulk to boundary propagators at the vertices enables performing computations at the level of truncated diagrams, hence the computations remain completely agnostic to what is attached from the boundary. This is no longer true for gravitons; thus, instead of working with the truncated graphs, we will work with the full Witten diagrams. Nonetheless, reducing the number of integrations to one by finding out an analogous relation for the graviton propagator will prove quite useful as we will see below.

To realize this goal, we first note the identity
\begin{multline}
\label{eq: decomposition of graviton tensor structure in terms of projectors}
\left(H^{(\omega,\bm{k})}_{ac}H^{(\omega,\bm{k})}_{bd}+H^{(\omega,\bm{k})}_{ad}H^{(\omega,\bm{k})}_{bc}-\frac{2}{d-1}H^{(\omega,\bm{k})}_{ab}H^{(\omega,\bm{k})}_{cd}\right)=\cP_{ab,cd}^{(1,1)\bm{k}}
\\+\frac{k^2+\omega^2}{\omega^2}\left(\cP_{ab,cd}^{(1,2)\bm{k}}+\cP_{ab,cd}^{(2,1)\bm{k}}\right)+\left(\frac{k^2+\omega^2}{\omega^2}\right)^2\cP_{ab,cd}^{(2,2)\bm{k}}\;,
\end{multline}
where we define
\be 
\cP_{ab,cd}^{(m,n)\bm{k}}\equiv\Pi^{(m)\bm{k}}_{ac}\Pi^{(n)\bm{k}}_{bd}+\Pi^{(m)\bm{k}}_{ad}\Pi^{(n)\bm{k}}_{bc}-\frac{2}{d-1}\Pi^{(m)\bm{k}}_{ab}\Pi^{(n)\bm{k}}_{cd}\;.
\ee 
{\color{red} }
This enables us to rewrite the graviton propagator in \equref{eq: graviton propagator} as
\be
\label{eq: graviton propagator explicit}
\mathcal{G}_{ab,cd}(\bm{k}; z, z')=&
\frac{i \cP_{ab,cd}^{(1,1)\bm{k}}}{2}\int
\omega d\omega 
\frac{(zz')^{\frac{d}{2}-2}J_{\nu}(\omega z) J_{\nu}(\omega z')}{k^2+\omega^2-i \epsilon}\\
&+
\frac{i\left(\cP_{ab,cd}^{(1,2)\bm{k}}+\cP_{ab,cd}^{(2,1)\bm{k}}+\cP_{ab,cd}^{(2,2)\bm{k}}\right)}{2}\int
\omega d\omega 
\frac{(zz')^{\frac{d}{2}-2} J_{\nu}(\omega z) J_{\nu}(\omega z')}{\omega^2}\\
&+
\frac{i\cP_{ab,cd}^{(2,2)\bm{k}}}{2}k^2\int
\omega d\omega
\frac{(zz')^{\frac{d}{2}-2} J_{\nu}(\omega z) J_{\nu}(\omega z')}{\omega^4}
\ee
hence we obtain the desired form
\be 
\mathcal{G}_{ab,cd}(\bm{k}; z, z')=\cD_{ab,cd}^{\bm{k}}\int\limits_{0}^{\infty}
\omega d\omega 
\frac{(zz')^{\frac{d}{2}-2} J_{\nu}(\omega z) J_{\nu}(\omega z')}{k^2+\omega^2-i \epsilon}\;,
\ee 
where we define the differential operator
\be
\label{eq: differential operator}
\cD_{ab,cd}^{\bm{k}}\coloneqq\frac{i}{2}\left[
\cP_{ab,cd}^{(0)\bm{k}}+\cP_{ab,cd}^{(1)\bm{k}}\lim\limits_{k\rightarrow 0}-\cP_{ab,cd}^{(2)\bm{k}}k^2\lim\limits_{k\rightarrow 0}\partial_{k^2}
\right]
\ee 
for
\be 
\cP_{ab,cd}^{(0)\bm{k}}\equiv\cP_{ab,cd}^{(1,1)\bm{k}}\;,\quad \cP_{ab,cd}^{(1)\bm{k}}\equiv\cP_{ab,cd}^{(1,2)\bm{k}}+\cP_{ab,cd}^{(2,1)\bm{k}}+\cP_{ab,cd}^{(2,2)\bm{k}}\;,\quad \cP_{ab,cd}^{(2)\bm{k}}\equiv  \cP_{ab,cd}^{(2,2)\bm{k}}\;.
\ee 

The nice thing about the operator $\cD^{\bm{k}}$ is that it commutes with the rest of the calculation, e.g. bulk point integration, so we can apply it at the very end. Furthermore, $\cD^{\bm{k}}$ for different propagators commute as well. Thus for a Witten diagram with $n$ bulk to bulk propagators we schematically have
\be 
\label{eq: schematic form}
\cA_{\text{Witten}}=\left(\epsilon_i,V_{i}\right)^{a_{11}a_{12}\dots a_{14}a_{21}\dots a_{n4}}\prod\limits_{j=1}^{n}\cD_{a_{j1}a_{j2},a_{j3}a_{j4}}^{\bm{p}_j}\cM\;,
\ee 
where $(\epsilon_i,V_i)$ stand for the collection of the vertex factors and polarization vectors, $\bm{p}_j$ is  sum of some bulk to boundary momenta depending on the topology of the diagram, and $\cM$ is the scalar factor of the amplitude: it is the graviton analog of the amplitude for the \emph{straight-only-graph} in \cite{Albayrak:2019asr} (and $\cM_{\text{topology}}^{(1)}$ analog of \cite{Albayrak:2018tam}). 
\subsection{Bulk point integrated expressions}

In the naive order of the calculations, one needs to 
carry out the integration in \equref{eq: graviton propagator explicit} to get the full propagator and then the radial integration to get the amplitude for the Witten diagram. The integration in the propagator comes from the equation of motion as derived in \cite{Raju:2011mp} and we need to carry out the bulk point integration because we go to the Fourier space of boundary coordinates only.

A clearer approach is to interchange the order of integrations as one needs same bulk-point integrated quantities for any tree level diagram; therefore, one can carry out bulk point integration once and for all as we did in our previous paper for gluons. In accordance with that paper's convention, we define  the bulk-point integrated objects $\mathcal{KKK}$, $\mathcal{KKJ}$, $\mathcal{KJJ}$, and $\mathcal{JJJ}$ as follows:
\small
\bea[eq: bulk point integrated] 
\mathcal{KKK}(k_1,k_2,k_3)\coloneqq&\int\limits_0^\infty \frac{dz}{z^{d+1}} z^8\left(\sqrt{\frac{2}{\pi}}z^{-2}(k_1 z)^{\frac{d}{2}} K_{\frac{d}{2}}(k_1 z)\right)\left(\sqrt{\frac{2}{\pi}}z^{-2}(k_2 z)^{\frac{d}{2}} K_{\frac{d}{2}}(k_2 z)\right)\nn\\&\x \left(\sqrt{\frac{2}{\pi}}z^{-2}(k_3 z)^{\frac{d}{2}} K_{\frac{d}{2}}(k_3 z)\right)\;,\\
\mathcal{KKJ}(k_1,k_2,k_3)\coloneqq&\int\limits_0^\infty \frac{dz}{z^{d+1}} z^8\left(\sqrt{\frac{2}{\pi}}z^{-2}(k_1 z)^{\frac{d}{2}} K_{\frac{d}{2}}(k_1 z)\right)\left(\sqrt{\frac{2}{\pi}}z^{-2}(k_2 z)^{\frac{d}{2}} K_{\frac{d}{2}}(k_2 z)\right)\nn\\&\x\left(z^{-2}( z)^{\frac{d}{2}} J_{\frac{d}{2}}(k_3 z)\right)\;,\\
\mathcal{KJJ}(k_1,k_2,k_3)\coloneqq&\int\limits_0^\infty \frac{dz}{z^{d+1}} z^8\left(\sqrt{\frac{2}{\pi}}z^{-2}(k_1 z)^{\frac{d}{2}} K_{\frac{d}{2}}(k_1 z)\right)\left(z^{-2}( z)^{\frac{d}{2}} J_{\frac{d}{2}}(k_2 z)\right)\left(z^{-2}( z)^{\frac{d}{2}} J_{\frac{d}{2}}(k_3 z)\right)\;,\\
\mathcal{JJJ}(k_1,k_2,k_3)\coloneqq&\int\limits_0^\infty \frac{dz}{z^{d+1}} z^8\left(z^{-2}( z)^{\frac{d}{2}} J_{\frac{d}{2}}(k_1 z)\right)\left(z^{-2}( z)^{\frac{d}{2}} J_{\frac{d}{2}}(k_2 z)\right)\left(z^{-2}( z)^{\frac{d}{2}} J_{\frac{d}{2}}(k_3 z)\right)\;,
\eea 
\normalsize
where $z^8$ factor comes from the contraction with the three point vertex $\cV^{ijklmn}_{\bm{k}_1, \bm{k}_2, \bm{k}_3}$. From now on, we will use only use the remaining tensor part $V^{ijklmn}_{\bm{k}_1, \bm{k}_2, \bm{k}_3}$ in the expressions.

In terms of these objects, the amplitude for a  generic tree level Witten diagram in \equref{eq: schematic form} takes the form\footnote{The factor $\mathcal{KKK}$ only appears in three point amplitude hence we ignored the dependence of $\cA_{\text{Witten}}$ on $\mathcal{KKK}$ in \equref{eq: generic form of the amplitude}. Likewise, the function $f$ may depend on other bulk point integrated objects, e.g. $\mathcal{KKJJ}$, if we allow interactions beyond cubic vertices.}
\be 
\label{eq: generic form of the amplitude}
\cA_{\text{Witten}}=\left(\epsilon_i,V_{i}\right)^{a_{11}a_{12}\dots a_{14}a_{21}\dots a_{n4}}\prod\limits_{j=1}^{n}\cD_{a_{j1}a_{j2},a_{j3}a_{j4}}^{\bm{p}_j}\int_0^\infty\prod\limits_{k=1}^n\frac{\omega_kd\omega_k}{p_k^2+\omega_k^2-i\epsilon}f(\mathcal{KKJ},\mathcal{KJJ},\mathcal{JJJ})\;,
\ee 
where the function $f$ depends on the topology of the diagram. Here $f(\mathcal{KKJ},\mathcal{KJJ},\mathcal{JJJ})$ depends on $k_1\cdots k_m$ and $\omega_1\cdots\omega_n$ where $m$ is the number of external legs and $n$ is the number of bulk to bulk propagators. Likewise, $\bm{p}_j$ and $p_k$ are combinations of external momenta depending on the topology.\footnote{Throughout the paper, we refer to expressions for particular Witten diagrams as \emph{amplitudes} for those diagrams for brevity. However, one should note that the full amplitudes are sums over various channels and relevant contact terms.}

As a non-trivial example, we can write the expression for the star-triangle six point diagram as
\be 
\cA_{\text{star-triangle}}=&
\left(\epsilon^1_{i_1j_1}\epsilon^2_{k_1l_1}
V^{i_1j_1k_1l_1m_1n_1}_{\bm{k}_1, \bm{k}_2, -\bm{k}_{\underline{12}}}\right)
\left(\epsilon^3_{i_2j_2}\epsilon^4_{k_2l_2}
V^{i_2j_2k_2l_2m_2n_2}_{\bm{k}_3, \bm{k}_4, -\bm{k}_{\underline{34}}}\right)
\left(\epsilon^5_{i_3j_3}\epsilon^6_{k_3l_3}
V^{i_3j_3k_3l_3m_3n_3}_{\bm{k}_5, \bm{k}_6, -\bm{k}_{\underline{56}}}\right)
\\&\hspace{-.6in}
\x V^{s_1t_1s_2t_2s_3t_3}_{\bm{k}_{\underline{12}},\bm{k}_{\underline{34}},\bm{k}_{\underline{56}}}
\cD_{m_1n_1s_1t_1}^{\bm{k}_{\underline{12}}}
\cD_{m_2n_2s_2t_2}^{\bm{k}_{\underline{34}}}
\cD_{m_3n_3s_3t_3}^{\bm{k}_{\underline{56}}}
\int_0^\infty\frac{\omega_1d\omega_1}{k_{\underline{12}}^2+\omega_1^2-i\epsilon}
\int_0^\infty\frac{\omega_2d\omega_2}{k_{\underline{34}}^2+\omega_2^2-i\epsilon}
\\&\hspace{-.6in}\x
\int_0^\infty\frac{\omega_3d\omega_3}{k_{\underline{56}}^2+\omega_3^2-i\epsilon}
\mathcal{KKJ}(k_1,k_2,\omega_1)
\mathcal{KKJ}(k_3,k_4,\omega_2)
\mathcal{KKJ}(k_5,k_6,\omega_3)
\mathcal{JJJ}(\omega_1,\omega_2,\omega_3)\;,
\ee 
where we are using the shorthand notation of \cite{Albayrak:2018tam}:
\be
\label{eq: notation}
k_{\underline{i_{11}i_{12}\dots i_{1n_1}}\;\underline{i_{21}i_{22}\dots i_{2n_2}}\dots \underline{i_{m1}i_{m2}\dots i_{mn_m}}j_1j_2\dots j_p}\coloneqq&\sum\limits_{a=1}^{m}\abs{\sum\limits_{b=1}^{n_a}\bm{k}_{i_{ab}}}+\sum\limits_{c=1}^{p}\abs{\bm{k}_{j_c}}\;,\\
\bm{k}_{i_1i_2\dots i_n}\coloneqq&\bm{k}_{i_1}+\bm{k}_{i_2}+\cdots+\bm{k}_{i_n}\;,
\ee
thus, e.g. $k_{123}=\abs{\bm{k}_1}+\abs{\bm{k}_2}+\abs{\bm{k}_3}\;,\;k_{1\underline{23}\;\underline{45}6}=\abs{\bm{k}_1}+\abs{\bm{k}_2+\bm{k}_3}+\abs{\bm{k}_4+\bm{k}_5}+\abs{\bm{k}_6}\;,\; \bm{k}_{12}=\bm{k}_1+\bm{k}_2$. Even though $\bm{k}_{12}$ can also be used to label $12^{\text{th}}$ momentum in a generic calculation, there will not be room for confusion in this paper.

In this paper, we are focusing on $d=3$, and we will need the explicit expressions for $\mathcal{KKJ}$ and $\mathcal{KJJ}$ only:
\bea 
\mathcal{KKJ}(k_1,k_2,k_3)={}&{}\frac{\sqrt{\frac{2}{\pi }} k_3^{3/2} \left(k_1^2+4 k_2 k_1+k_2^2+k_3^2\right)}{\left(\left(k_1+k_2\right)^2+k_3^2\right)^2}\;,\\
\mathcal{KJJ}(k_1,k_2,k_3)={}&{}\frac{32 k_1^3 \left(k_2 k_3\right)^{3/2}}{\pi  \left(k_1^4+2 \left(k_2^2+k_3^2\right) k_1^2+\left(k_2^2-k_3^2\right)^2\right)^2}\;.
\eea 
\section{Amplitudes}
\label{sec:3}
\subsection{A basic Witten diagram: three point function} 
\label{sec:threepoint}
\begin{figure} 
	\centering
	\includegraphics[width=.4\textwidth,origin=c]{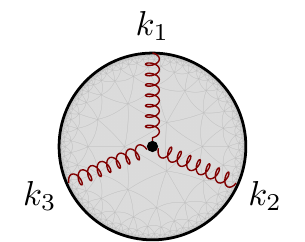}
	\caption{The three point graviton amplitude \label{fig: 3pt}}
\end{figure}
As a warm-up, we will compute the three point amplitude. It is a useful exercise as all the dynamical data in a CFT is actually contained solely in three point correlation functions \cite{Bzowski:2015pba, Bzowski:2017poo}.\footnote{We will  analyze three point function dual to CFT correlator of stress tensors only; to access full dynamical data, one of course needs to consider three point function of all multitrace operators built out of the stress tensor.} With the ingredients in \equref{eq: graviton ingredients}, we can calculate the momentum space  Witten diagram shown in \figref{\ref{fig: 3pt}}:
\be
\cA_{3}=\int\limits_{0}^{\infty}\frac{dz}{z^{d+1}}h_{ij}(\bm{k}_1,z)h_{kl}(\bm{k}_2,z)h_{mn}(\bm{k}_3,z)\cV^{ijklmn}_{\bm{k}_1, \bm{k}_2, \bm{k}_3}
\ee
which then reads
\be
\cA_{3}=
\epsilon_{ij}^1\epsilon_{kl}^2\epsilon_{mn}^3V^{ijklmn}_{\bm{k}_1, \bm{k}_2, \bm{k}_3}\left(\frac{2}{\pi}\right)^{3/2}(k_1k_2k_3)^{\frac{d}{2}}
\int\limits_{0}^{\infty}z^{\frac{d+2}{2}}dzK_{\frac{d}{2}}(k_1 z)K_{\frac{d}{2}}(k_2 z)K_{\frac{d}{2}}(k_3 z)\;.
\ee
In $d=3$, the integration becomes
\begin{multline}
\left(\frac{2}{\pi}k_1k_2k_3\right)^{3/2}\int\limits z^{\frac{5}{2}}dzK_{\frac{3}{2}}(k_1 z)K_{\frac{3}{2}}(k_2 z)K_{\frac{3}{2}}(k_3 z)= -\frac{k_1 k_2 k_3 z e^{\left(-k_1-k_2-k_3\right) z}}{k_1+k_2+k_3}\\-\frac{k_1 k_2 k_3
	e^{\left(-k_1-k_2-k_3\right) z}}{\left(k_1+k_2+k_3\right)^2}-\frac{\left(k_2 k_3+k_1
	k_2+k_1 k_3\right) e^{\left(-k_1-k_2-k_3\right)
		z}}{k_1+k_2+k_3}-\frac{e^{\left(-k_1-k_2-k_3\right) z}}{z}
\end{multline}
which yields the regularized result
\be
\cA_{3}=
\epsilon_{ij}^1\epsilon_{kl}^2\epsilon_{mn}^3V^{ijklmn}_{\bm{k}_1, \bm{k}_2, \bm{k}_3}
\left[
\frac{k_1 k_2 k_3}{\left(k_1+k_2+k_3\right)^2}+\frac{k_2 k_3+k_1
	k_2+k_1 k_3}{k_1+k_2+k_3}+\left(k_1+k_2+k_3\right)
\right]\;.
\ee
This result also appears in \cite{Maldacena:2011nz} where instead of $z$ integral the details appears from the computation of the time integral. Finally, we also know that,
\be 
\label{eq: three point explicit}
\epsilon_{ij}^1\epsilon_{kl}^2\epsilon_{mn}^3V^{ijklmn}_{\bm{k}_1, \bm{k}_2, \bm{k}_3}={}&{} \epsilon_{ij}^1\epsilon_{kl}^2\epsilon_{mn}^3\left(
\frac{(\bm{k}_2)^i(\bm{k}_3)^j\eta^{km}\eta^{ln}}{4}-\frac{(\bm{k}_2)^i(\bm{k}_3)^k\eta^{jm}\eta^{ln}}{2}
\right)\\&+\epsilon_{ij}^1\epsilon_{kl}^2\epsilon_{mn}^3\left(
\frac{(\bm{k}_3)^k(\bm{k}_1)^l\eta^{mi}\eta^{nj}}{4}-\frac{(\bm{k}_3)^k(\bm{k}_1)^m\eta^{li}\eta^{nj}}{2}
\right)
\\&+\epsilon_{ij}^1\epsilon_{kl}^2\epsilon_{mn}^3\left(
\frac{(\bm{k}_1)^m(\bm{k}_2)^n\eta^{ik}\eta^{jl}}{4}-\frac{(\bm{k}_1)^m(\bm{k}_2)^i\eta^{nk}\eta^{jl}}{2}
\right)
\\
=&{}\frac{1}{4}\left[\left(\bm{k}_2\epsilon_1\bm{k}_3\right)(\epsilon_2\epsilon_3)+\left(\bm{k}_3\epsilon_2\bm{k}_1\right)(\epsilon_3\epsilon_1)+\left(\bm{k}_1\epsilon_3\bm{k}_2\right)(\epsilon_1\epsilon_2)\right]\\&{}-
\frac{1}{2}\left[\left(\bm{k}_2\epsilon_1\epsilon_3\epsilon_2\bm{k}_3\right)+\left(\bm{k}_3\epsilon_2\epsilon_1\epsilon_3\bm{k}_1\right)+\left(\bm{k}_1\epsilon_3\epsilon_2\epsilon_1\bm{k}_2\right)\right]
\ee 
with the shorthand notation $(\bm{k}_a\epsilon_{b_1}\dots\epsilon_{b_n}\bm{k}_b)\equiv(\bm{k}_a)^{i_1}\epsilon^{b_1}_{i_1i_2}\eta^{i_2i_3}\epsilon^{b_2}_{i_3i_4}\eta^{i_4i_5}\dots\epsilon^{b_n}_{i_{2n-1}i_{2n}}(\bm{k}_a)^{i_{2n}}$ and $(\epsilon_a\epsilon_b)\equiv \epsilon^{a}_{ij}\epsilon^{b}_{kl}\eta^{ik}\eta^{jl}$.

Therefore, the amplitude associated with the three point diagram reads explicitly as
\be
\label{eq: three point amplitude}
\cA_{3}=&{}\bigg[\frac{1}{4}\left(\left(\bm{k}_2\epsilon_1\bm{k}_3\right)(\epsilon_2\epsilon_3)+\left(\bm{k}_3\epsilon_2\bm{k}_1\right)(\epsilon_3\epsilon_1)+\left(\bm{k}_1\epsilon_3\bm{k}_2\right)(\epsilon_1\epsilon_2)\right)\\&{}-
\frac{1}{2}\left(\left(\bm{k}_2\epsilon_1\epsilon_3\epsilon_2\bm{k}_3\right)+\left(\bm{k}_3\epsilon_2\epsilon_1\epsilon_3\bm{k}_1\right)+\left(\bm{k}_1\epsilon_3\epsilon_2\epsilon_1\bm{k}_2\right)\right)
\bigg]
\\&{}\x\left[
\frac{k_1 k_2 k_3}{\left(k_1+k_2+k_3\right)^2}+\frac{k_2 k_3+k_1
	k_2+k_1 k_3}{k_1+k_2+k_3}+\left(k_1+k_2+k_3\right)
\right]\;.
\ee

\subsection{Four point function}
\begin{figure} 
	\centering
	\includegraphics[width=.4\textwidth,origin=c]{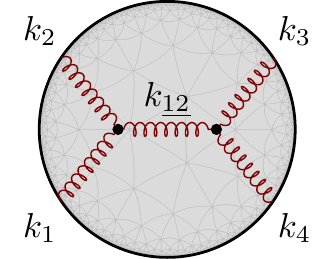}
	\caption{The four point graviton amplitude \label{fig: 4pt}}
\end{figure}
In this section, we will step by step compute the amplitude associated with the $s-$channel four point Witten diagram shown in \figref{\ref{fig: 4pt}}.\footnote{One can write down expressions associated with other channels in a similar fashion.} For this topology, \equref{eq: generic form of the amplitude} becomes
\begin{multline}
\label{eq: four point amplitude}
\cA_{\text{four point}}=
\left(\epsilon^1_{i_1j_1}\epsilon^2_{k_1l_1}
V^{i_1j_1k_1l_1m_1n_1}_{\bm{k}_1, \bm{k}_2, \bm{k}_{\underline{12}}}\right)
\left(\epsilon^3_{i_2j_2}\epsilon^4_{k_2l_2}
V^{i_2j_2k_2l_2m_2n_2}_{\bm{k}_3, \bm{k}_4, -\bm{k}_{\underline{12}}}\right)
\\\x\cD_{m_1n_1m_2n_2}^{\bm{k}_{\underline{12}}}
\int_0^\infty\frac{\omega d\omega}{k_{\underline{12}}^2+\omega^2-i\epsilon}
\mathcal{KKJ}(k_1,k_2,\omega)
\mathcal{KKJ}(k_3,k_4,\omega)\;.
\end{multline}

We first focus on the scalar factor. From \equref{eq: bulk point integrated}, it reads as
\be 
\mathcal{M}_{\text{four point}}\coloneqq&\int_0^\infty\frac{\omega d\omega}{k_{\underline{12}}^2+\omega^2-i\epsilon}
\mathcal{KKJ}(k_1,k_2,\omega)
\mathcal{KKJ}(k_3,k_4,\omega)
\\
=& \int_0^\infty d\omega  \frac{2 \omega ^4 \left(k_1^2+k_2^2+4 k_1 k_2+\omega ^2\right) \left(k_3^2+k_4^2+4 k_3 k_4+\omega ^2\right)}{\pi 
	\left(\left(k_1+k_2\right){}^2+\omega ^2\right){}^2 \left(\left(k_3+k_4\right){}^2+\omega ^2\right){}^2 \left(k_{\underline{12}}^2+\omega
	^2-i\epsilon\right)}
\ee 
We can carry out this integration\footnote{As advocated in \cite{Albayrak:2018tam}, we use residue theorem to efficiently compute these symbolic integrations. We numerically verified this result, as well as the other symbolic integrations we compute below.} to find 
\scriptsize
\be 
\mathcal{M}_{\text{four point}}=&\frac{1}{\left(-k_{\underline{12}}+k_1+k_2\right)^2
	\left(k_{\underline{12}}+k_1+k_2\right)^2}\bigg(
\frac{2 k_1 k_2 k_3 \left(k_1+k_2+k_3\right) \left(-k_{\underline{12}}+k_1+k_2\right)
		\left(k_{\underline{12}}+k_1+k_2\right)}{\left(k_1+k_2+k_3+k_4\right)^3}\\&-\frac{k_3 k_{\underline{12}} \left(k_{\underline{12}}+k_3\right)
		\left(-k_{\underline{12}}^2+k_1^2+4 k_2
		k_1+k_2^2\right)}{\left(k_{\underline{12}}+k_3+k_4\right)^2}-\frac{\left(-k_{\underline{12}}^2+k_1^2+4 k_2 k_1+k_2^2\right)
		\left(k_{\underline{12}}^2+k_3^2\right)}{k_{\underline{12}}+k_3+k_4}\\&+\frac{k_1^2 \left(-k_{\underline{12}}^2+6
		k_2^2+k_3^2\right)-\left(k_2^2+k_3^2\right) \left(k_{\underline{12}}-k_2\right) \left(k_{\underline{12}}+k_2\right)+k_1^4+4 k_2 k_1^3+4 k_2
		\left(k_2^2+k_3^2\right) k_1}{k_1+k_2+k_3+k_4}\\&+\frac{\left(k_1+k_2\right) k_3^2 \left(-k_{\underline{12}}^2+k_1^2+4 k_2 k_1+k_2^2\right)+k_3
		\left(-k_1^2 \left(k_{\underline{12}}^2-6 k_2^2\right)-k_2^2 k_{\underline{12}}^2+k_1^4+4 k_2 k_1^3+4 k_2^3 k_1+k_2^4\right)}{\left(k_1+k_2+k_3+k_4\right)^2}\\&+\frac{k_1 k_2
		\left(k_1+k_2\right) \left(-k_{\underline{12}}+k_1+k_2\right)
		\left(k_{\underline{12}}+k_1+k_2\right)}{\left(k_1+k_2+k_3+k_4\right)^2}\bigg)\;.
\ee 
\normalsize
The result seems rather complicated; however it can be put in a simpler and more intuitive form. For this, we need to make use of the symmetries of the four point exchange diagram. As we can see from \figref{\ref{fig: 4pt}}, the scalar factor $\mathcal{M}_{\text{four point}}$ should have the symmetries $k_1\leftrightarrow k_2$, $k_3\leftrightarrow k_4,$ and $\{k_1,k_2\}\leftrightarrow \{k_3,k_4\}$, hence we can change variables to those invariant under these interchanges:
\be 
\label{eq: four point parameters}
E\equiv&k_1+k_2+k_3+k_4\;,\\
\a\equiv&(k_1+k_2)(k_3+k_4)\;,\\
\b\equiv&k_1k_2+k_3k_4\;,\\
\g\equiv&k_1k_2k_3k_4\;,\\
\s\equiv&\abs{k_1+k_2-k_3-k_4}\;,\\
\lambda\equiv&\abs{k_1k_2-k_3k_4}\;.
\ee 
Of course, not all of these parameters are independent; in fact, we have the relations \mbox{$E^2-4\a=\s^2$} and \mbox{$\b^2-4\g=\lambda^2$}. With these parameters, the scalar factor now reads as
\begin{multline}
\mathcal{M}_{\text{four point}}=\frac{1}{\left(E k_{\underline{12}}+k_{\underline{12}}^2+\alpha \right)^2}\bigg(\frac{2 \gamma  \left(k_{\underline{12}}^2+\alpha \right)}{E^3}-\frac{-8 \gamma  k_{\underline{12}}+\lambda  \sigma  k_{\underline{12}}^2+\alpha 
	\lambda  \sigma }{2 E^2}\\+\frac{k_{\underline{12}}^2 (2 \alpha +3 \beta )-2 \lambda  \sigma  k_{\underline{12}}+\alpha  (2 \alpha +\beta )}{2
	E}+E k_{\underline{12}}^2+k_{\underline{12}} \left(k_{\underline{12}}^2+2 \alpha +\beta \right)\bigg)\;.
\end{multline} 
By acting our differential operator in \equref{eq: differential operator}, 
\footnotesize
\be 
\label{eq: four point full tensor}
{}&\mathcal{M}^{\text{four point}}_{m_1n_1m_2n_2}\coloneqq-2i\cD_{m_1n_1m_2n_2}^{\bm{k}_{\underline{12}}}\mathcal{M}_{\text{four point}}=\\
{}&
\frac{\frac{\lambda  \sigma  \left(k_{\underline{12}}^2 \mathcal{P}_{m_1 n_1,m_2 n_2}^{(2) k_{\underline{12}}}-\alpha  \left(\frac{\alpha 
			\left(k_{\underline{12}}^2+\alpha \right) \mathcal{P}_{m_1 n_1,m_2 n_2}^{(0) k_{\underline{12}}}}{\left(k_{\underline{12}}
			\left(k_{\underline{12}}+E\right)+\alpha \right){}^2}+\mathcal{P}_{m_1 n_1,m_2 n_2}^{(1) k_{\underline{12}}}\right)\right)}{2 \alpha ^2}+\frac{4
		\gamma  k_{\underline{12}} \mathcal{P}_{m_1 n_1,m_2 n_2}^{(0) k_{\underline{12}}}}{\left(k_{\underline{12}}
		\left(k_{\underline{12}}+E\right)+\alpha \right)^2}}{E^2}
\\&+
\frac{k_{\underline{12}} \left(k_{\underline{12}}^2+2 \alpha +\beta \right) \mathcal{P}_{m_1 n_1, m_2 n_2}^{(0)
		k_{\underline{12}}}}{\left(k_{\underline{12}} \left(k_{\underline{12}}+E\right)+\alpha \right)^2}+E\left(\frac{k_{\underline{12}}^2 \mathcal{P}_{m_1 n_1,m_2 n_2}^{(0) k_{\underline{12}}}}{\left(k_{\underline{12}} \left(k_{\underline{12}}+E\right)+\alpha
	\right){}^2}-\frac{\beta  k_{\underline{12}}^2 \mathcal{P}_{m_1 n_1,m_2 n_2}^{(2) k_{\underline{12}}}}{2 \alpha ^3}\right)
\\&+\frac{2 \gamma  \left(\alpha  \left(\frac{\alpha  \left(k_{\underline{12}}^2+\alpha \right) \mathcal{P}_{m_1 n_1,m_2 n_2}^{(0)
			k_{\underline{12}}}}{\left(k_{\underline{12}} \left(k_{\underline{12}}+E\right)+\alpha \right){}^2}+\mathcal{P}_{m_1 n_1,m_2 n_2}^{(1)
		k_{\underline{12}}}\right)-k_{\underline{12}}^2 \mathcal{P}_{m_1 n_1,m_2 n_2}^{(2) k_{\underline{12}}}\right)}{\alpha ^2 E^3}+\frac{\lambda  \sigma  k_{\underline{12}}^2 \mathcal{P}_{m_1 n_1,m_2 n_2}^{(2) k_{\underline{12}}}}{2 \alpha ^3}\\
	&+\frac{\alpha ^2 \left(\frac{\alpha  \left(k_{\underline{12}}^2 (2 \alpha +3 \beta )-2 \lambda  \sigma  k_{\underline{12}}+\alpha  (2 \alpha +\beta
			)\right) \mathcal{P}_{m_1 n_1,m_2 n_2}^{(0) k_{\underline{12}}}}{\left(k_{\underline{12}} \left(k_{\underline{12}}+E\right)+\alpha
			\right){}^2}+(2 \alpha +\beta ) \mathcal{P}_{m_1 n_1,m_2 n_2}^{(1) k_{\underline{12}}}\right)+k_{\underline{12}}^2 (\alpha  (\beta -2 \alpha )-4
		\gamma ) \mathcal{P}_{m_1 n_1,m_2 n_2}^{(2) k_{\underline{12}}}}{2 \alpha ^3 E}\;,
\ee 
\normalsize
hence \equref{eq: four point amplitude} reads as
\begin{equation}
\label{eq: four point result}
\cA_{\text{four point}}=\frac{i}{2}
\left(\epsilon^1_{i_1j_1}\epsilon^2_{k_1l_1}
V^{i_1j_1k_1l_1m_1n_1}_{\bm{k}_1, \bm{k}_2, \bm{k}_{\underline{12}}}\right)
\mathcal{M}^{\text{four point}}_{m_1n_1m_2n_2}
\left(\epsilon^3_{i_2j_2}\epsilon^4_{k_2l_2}
V^{i_2j_2k_2l_2m_2n_2}_{\bm{k}_3, \bm{k}_4, -\bm{k}_{\underline{12}}}\right)\;.
\end{equation}

\subsection{Five point function}
\begin{figure} 
	\centering
	\includegraphics[width=.3\textwidth,origin=c]{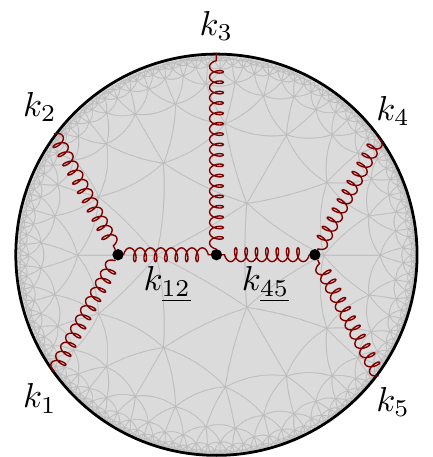}
	\caption{Five point graviton amplitude \label{fig: 5pt}}
\end{figure}
Here we will compute the amplitude associated with the five point Witten diagram shown in \figref{\ref{fig: 5pt}}. Alongside being an interesting computational challange as it contains two bulk to bulk propagators, this Comb channel is the only diagram one can write without studying beyond cubic interactions. 

Analogous to the four point diagram, we can write down the five point amplitude as
\begin{multline}
\label{eq: five point amplitude}
\cA_{\text{five point}}=
\left(\epsilon^1_{i_1j_1}\epsilon^2_{k_1l_1}
V^{i_1j_1k_1l_1m_1n_1}_{\bm{k}_1, \bm{k}_2, \bm{k}_{\underline{12}}}\right)
\left(\epsilon^3_{i_3j_3}
V^{p_1r_1i_3j_3p_2r_2}_{-\bm{k}_{\underline{12}}, \bm{k}_3, -\bm{k}_{\underline{45}}}\right)
\left(\epsilon^4_{i_2j_2}\epsilon^5_{k_2l_2}
V^{i_2j_2k_2l_2m_2n_2}_{\bm{k}_4, \bm{k}_5, \bm{k}_{\underline{45}}}\right)
\\\x\cD_{m_1n_1p_1r_1}^{\bm{k}_{\underline{12}}}\cD_{m_2n_2p_2r_2}^{\bm{k}_{\underline{45}}}\cM_{\text{five point}}\;,
\end{multline}
where we define the five point scalar factor
\begin{multline}
\cM_{\text{five point}}\equiv\int_0^\infty\frac{\omega_1\omega_2 d\omega_1d\omega_2}{\left(k_{\underline{12}}^2+\omega_1^2-i\epsilon\right)\left(k_{\underline{45}}^2+\omega_2^2-i\epsilon\right)}\\
\x
\mathcal{KKJ}(k_1,k_2,\omega_1)
\mathcal{KJJ}(k_3,\omega_1,\omega_2)\mathcal{KKJ}(k_4,k_5,\omega_2)\;.
\end{multline}
Such scalar factors can be calculated in a symbolic calculation program such as \texttt{Mathematica}, especially by taking advantage of the residue theorem as we demonstrated in our previous work. The result takes the following nice form
\be 
\cM_{\text{five point}}={}&{}\frac{\sum\limits_{i=0}^{9}k_3^i\left(c_i+(k_{\underline{12}}-k_{\underline{45}})d_i\right)}{2k_{12\underline{12}}^2k_{45\underline{45}}^2k_{\underline{12}3\underline{45}}^2k_{123\underline{45}}^3k_{\underline{12}345}^3k_{12345}^4}\;.
\ee 
By using an appropriate set of variables analogous to \equref{eq: four point parameters} we can write the coefficients $c_i, d_i$ in a simpler form; however, they are still rather complicated so we list the explicit expressions in Appendix~\ref{appendix}. The reader can also refer to the attached Mathematica file where the result is presented in terms of the conventional variables $k_i$.

With $\cM_{\text{five point}}$, one can straightforwardly obtain $\cA_{\text{five point}}$ as we have demonstrated above for four point amplitude. Even though the result looks quite complicated, it can simplify in certain limits and the commutation of the limiting procedure with the act of differential operators is one of the main strengths of our formalism since it allows us to obtain final elegantly simple results in a relatively easy computation. We will discuss two such limits in the next section.

\section{Flat space and collinear limits}
\label{sec: limits}
\paragraph{Flat space limit:} As mentioned in the introduction, it has been known that Witten diagrams reduce to scattering amplitudes in an appropriate flat space limit  \cite{Raju:2012zr,Raju:2012zs,Maldacena:2011nz,Arkani-Hamed:2015bza}. One of the advantages of using momentum space is that this procedure becomes almost trivial. In \cite{Raju:2012zr}, the author derives the relation that extracts the flat space amplitude from the AdS correlators. For tree-level graviton diagrams in AdS$_4$, the relation simply reads as
\be 
\label{eq: generic flat space limit}
\cA^{\text{flat space}}_{n-\text{point}}=\lim\limits_{E\rightarrow0}\frac{E^{n-1}}{\Gamma(n-1)\prod_{i=1}^nk_i}\cA_{n-\text{point}}
\ee 
for $E=k_1+\dots+k_n$.

Let us apply this formula for the amplitudes that we examined above. In the case of three point diagram, we can immediately read off the result from \equref{eq: three point amplitude}:
\be
\cA_{3}^{\text{flat space}}=&{}\bigg[\frac{1}{4}\left(\left(\bm{k}_2\epsilon_1\bm{k}_3\right)(\epsilon_2\epsilon_3)+\left(\bm{k}_3\epsilon_2\bm{k}_1\right)(\epsilon_3\epsilon_1)+\left(\bm{k}_1\epsilon_3\bm{k}_2\right)(\epsilon_1\epsilon_2)\right)\\&{}-
\frac{1}{2}\left(\left(\bm{k}_2\epsilon_1\epsilon_3\epsilon_2\bm{k}_3\right)+\left(\bm{k}_3\epsilon_2\epsilon_1\epsilon_3\bm{k}_1\right)+\left(\bm{k}_1\epsilon_3\epsilon_2\epsilon_1\bm{k}_2\right)\right)
\bigg]\;.
\ee
The transition to flat space limit is also quite transparent for four point diagrams. In particular, in the parameters that we introduced in \equref{eq: four point parameters}, \equref{eq: generic flat space limit} becomes
\be 
\cA_{\text{four point}}^{\text{flat space}}=\lim\limits_{E\rightarrow 0}\frac{E^3}{2\gamma}\cA_{\text{four point}}\;.
\ee 
Inserting \equref{eq: four point result} into this, we obtain the simple result
\begin{multline}
\cA_{\text{four point}}^{\text{flat space}}=\frac{i}{2}
\left(\epsilon^1_{i_1j_1}\epsilon^2_{k_1l_1}
V^{i_1j_1k_1l_1m_1n_1}_{\bm{k}_1, \bm{k}_2, \bm{k}_{\underline{12}}}\right)
\left(\epsilon^3_{i_2j_2}\epsilon^4_{k_2l_2}
V^{i_2j_2k_2l_2m_2n_2}_{\bm{k}_3, \bm{k}_4, -\bm{k}_{\underline{12}}}\right)
\\\x\left(\frac{\mathcal{P}_{m_1 n_1,m_2 n_2}^{(0) k_{\underline{12}}}}{k_{\underline{12}}^2+\alpha }+\frac{\mathcal{P}_{m_1 n_1,m_2 n_2}^{(1)
		k_{\underline{12}}}}{\alpha }-\frac{k_{\underline{12}}^2 \mathcal{P}_{m_1 n_1,m_2 n_2}^{(2) k_{\underline{12}}}}{\alpha ^2}\right)\;.
\end{multline}
By inserting the explicit expressions for $\cP$ and $V$, one can reduce the result to product of momenta and the polarization vectors; for example,
\footnotesize
\be 
{}&\left(\epsilon^1_{i_1j_1}\epsilon^2_{k_1l_1}
V^{i_1j_1k_1l_1m_1n_1}_{\bm{k}_1, \bm{k}_2, \bm{k}_{\underline{12}}}\right)
\mathcal{P}_{m_1 n_1,m_2 n_2}^{(2) k_{\underline{12}}}
\left(\epsilon^3_{i_2j_2}\epsilon^4_{k_2l_2}
V^{i_2j_2k_2l_2m_2n_2}_{\bm{k}_3, \bm{k}_4, -\bm{k}_{\underline{12}}}\right)=-\frac{1}{16k_{\underline{12}}^4}\\&\x\left[
(\bm{k}_{\underline{12}}\epsilon_1\bm{k}_{\underline{12}})(\bm{k}_{\underline{12}}\epsilon_2\bm{k}_1)-(\bm{k}_{\underline{12}}\epsilon_2\bm{k}_{\underline{12}})(\bm{k}_{\underline{12}}\epsilon_1\bm{k}_2)+\bm{k}_{1}\.\bm{k}_{\underline{12}}\left(
\bm{k}_{2}\.\bm{k}_{\underline{12}}(\epsilon_1\epsilon_2)-2(\bm{k}_{\underline{12}}\epsilon_1\epsilon_2\bm{k}_{\underline{12}})-2(\bm{k}_{2}\epsilon_1\epsilon_2\bm{k}_{\underline{12}})
\right)
\right]\\
&\x\left[
(\bm{k}_{\underline{12}}\epsilon_4\bm{k}_{\underline{12}})(\bm{k}_{\underline{12}}\epsilon_3\bm{k}_4)-(\bm{k}_{\underline{12}}\epsilon_3\bm{k}_{\underline{12}})(\bm{k}_{\underline{12}}\epsilon_4\bm{k}_3)+\bm{k}_{3}\.\bm{k}_{\underline{12}}\left(
\bm{k}_{4}\.\bm{k}_{\underline{12}}(\epsilon_3\epsilon_4)+2(\bm{k}_{\underline{12}}\epsilon_3\epsilon_4\bm{k}_{\underline{12}})-2(\bm{k}_{4}\epsilon_3\epsilon_4\bm{k}_{\underline{12}})
\right)
\right]
\ee 
\normalsize
with the shorthand notation used in \equref{eq: three point explicit}.

We can similarly calculate the flat space limit for five point diagram. For brevity, we will only provide the flat space limit of $\cM_{\text{five point}}$; one can obtain the full flat space amplitude $\cA_{\text{five point}}$ by acting with the differential operators and carrying out the relevant contractions as we demonstrated above for the four point case. We emphasize that the flat space limit commutes with the differential operators, thus this procedure is valid. The result simply reads as
\be 
\cM_{\text{five point}}^{\text{flat space}}=\lim\limits_{k_{12345}\rightarrow 0} \frac{k_{12345}^4}{6k_1k_2k_3k_4k_5}\cM_{\text{five point}}={}\frac{1}{(k_{\underline{12}}^2-k_{12}^2)(k_{\underline{123}}^2-k_{123}^2)}\;.
\ee 

At the level of scalar factors, the flat space limit for four point function takes a similar form
\be 
\cM_{\text{four point}}^{\text{flat space}}=\lim\limits_{k_{1234}\rightarrow 0} \frac{k_{1234}^3}{2k_1k_2k_3k_4}\cM_{\text{four point}}={}&{}\frac{1}{k_{\underline{12}}^2-k_{12}^2}\;.
\ee 
In fact, looking at these results, we may generalize the flat space limit of scalar factors for higher point comb-like diagrams as
\be 
\cM_{n-\text{point comb-like diagram}}^{\text{flat space}}=\frac{1}{\prod\limits_{i=1}^{n-2}
	\left(\abs{\sum\limits_{j=1}^i\bm{k}_j}^2-\left(\sum\limits_{j=1}^i\abs{\bm{k}_j}\right)^2\right)}\;.
\ee

\paragraph{Collinear limit:}
Another interesting limit that we may manifestly take in our formalism is the collinear limit, e.g. when vector sum of the momenta of two external legs approaches zero. In most of the cases, our relations do not provide an immediate simplification albeit they are perfectly suitable for the calculation. However, in the cases where the chosen external legs directly interact in the chosen Witten diagram, we may drastically simplify the calculation by commuting the collinear limit with the differential operator for the relevant bulk to bulk propagator.

We can see this simplification as follows. Let us consider the collinear limit of a generic Witten diagram
\be 
\lim\limits_{\bm{q}\rightarrow 0}\cA_{\text{Witten}}=\left[\lim\limits_{\bm{q}\rightarrow 0}\left(\epsilon_i,V_{i}\right)^{a_{11}a_{12}\dots a_{14}a_{21}\dots a_{n4}}\right]
\lim\limits_{\bm{q}\rightarrow 0}\prod\limits_{j=1}^{n}\cD_{a_{j1}a_{j2},a_{j3}a_{j4}}^{\bm{p}_j}\cM\;,
\ee 
where $\bm{q}$ is the vanishing momentum in the chosen collinear limit, $\bm{p}_j$ is sum of some bulk to boundary momenta depending on the topology of the diagram and $\cM$ is the scalar part of the amplitude. If $\bq\in\{\bp_j\}$, taking $\bp_n=\bq$,
we see that
\begin{multline}
\lim\limits_{\bm{q}\rightarrow 0}\cA_{\text{Witten}}=-\frac{i}{2}\left(\eta_{a_{n1}a_{n3}}\eta_{a_{n2}a_{n4}}-\eta_{a_{n1}a_{n2}}\eta_{a_{n3}a_{n4}}+\eta_{a_{n1}a_{n4}}\eta_{a_{n2}a_{n3}}\right)
\\\x\left[\lim\limits_{\bm{q}\rightarrow 0}\left(\epsilon_i,V_{i}\right)^{a_{11}a_{12}\dots a_{14}a_{21}\dots a_{n4}}\right] \prod\limits_{j=1}^{n-1}\cD_{a_{j1}a_{j2},a_{j3}a_{j4}}^{\bm{p}_j}\left(\lim\limits_{\bm{q}\rightarrow0}\cM\right)\;.
\end{multline}
This follows from the identity
\be 
\lim\limits_{\bk\rightarrow 0} \cD_{ab,cd}^{\bm{k}}\cM=\frac{\eta_{ac}\eta_{bd}-\eta_{ab}\eta_{bd}+\eta_{ad}\eta_{bc}}{2i}\lim\limits_{\bk\rightarrow 0}\cM\;,
\ee 
which can be easily verified with \equref{eq: differential operator}.

We can see this at work by analyzing the collinear limit $\bk_{12}\rightarrow 0$ for the four point amplitude in \equref{eq: four point amplitude} where the drastic simplification in \equref{eq: four point full tensor} reads as
\be 
\lim\limits_{\bm{k}_{\underline{12}}\rightarrow 0}\mathcal{M}^{\text{four point}}_{m_1n_1m_2n_2}=-
\frac{\left(4 \gamma +2 \alpha  E^2+\beta  E^2-E \lambda  \sigma \right)}{2 \alpha  E^3}\left(\eta_{m_1m_2}\eta_{n_1n_2}-\eta_{m_1n_1}\eta_{m_2n_2}+\eta_{m_1n_2}\eta_{m_2n_1}\right)\;.
\ee
This leads to
\be 
\lim\limits_{\bm{k}_{\underline{12}}\rightarrow 0}\cA_{\text{four point}}=i\frac{41k^3}{4096}
\cos(2\theta)(\epsilon_1\epsilon_2)(\epsilon_3\epsilon_4)\;,
\ee 
where we took $\{E,\a,\b,\g,\sigma,\lambda\}\rightarrow\{4k,4k^2,2k^2,k^4,0,0\}$ and defined $\bm{k}_1\.\bm{k}_3=k^2\cos(\theta)$ for $\abs{\bm{k}_1}=\abs{\bm{k}_3}=k$.

We note that flat space limit and collinear limit do not commute! For instance, taking the flat space limit \emph{after} the collinear limit kills the four point amplitude whereas we get a finite result in the reverse order:
\be 
\lim\limits_{\bm{k}_{\underline{12}}\rightarrow 0}\cA_{\text{four point}}^{\text{flat space}}=i\frac{k^2}{128}
\cos(2\theta)(\epsilon_1\epsilon_2)(\epsilon_3\epsilon_4)\;.
\ee 

\section{Conclusion}

In this paper, we have studied graviton propagating in AdS$_4$. As a continuation of our previous work with the gluons, we aimed to adapt and further develop the existent momentum space technology to graviton interactions, and to provide explicit results for tree level calculations. For this, we rewrote the graviton propagator in a way that reduced the complexity of the problem significantly. Explicitly, we managed to reduce the bulky part of the computation for any diagram to the calculation of a scalar factor. This rearranged form
of the amplitudes may lead to a more systematic and overarching study of AdS correlators in momentum space. For instance, it is conceivable that these objects have similar recursive structures as the gauge theory correlators computed in \cite{Albayrak:2019asr}. Finding an analogous algorithmic method to compute the graviton interactions is an open problem which we hope to address in the future.

Our formalism has also practical importance in addition to being a possible step to a more systematic study of momentum space AdS correlators. In a standard calculation, one needs to carry out bulk point integrations whose number increases exponentially with the number of bulk to bulk propagators. In our settings, there is only one scalar factor per diagram, hence we need to carry out only one integration for any diagram.\footnote{Technically, there are $3^n$ multiple integrals of $n$ variables in standard approach where $n$ is the number of bulk to bulk propagators, and there is only one multiple integral of $n$ variables in our approach. Not all $3^n$ of these integrations are independent, and in fact one can reduce the number of integrations significantly by using the symmetries of the diagram. Nonetheless, we bypass any such additional analysis and reduce the number for any diagram.} As the integrands consist of several Bessel functions, this is a rather important simplification which enables the calculations of higher point functions in practice. We demonstrated this utility by computing four and five point Witten diagrams. While these expressions are naively complicated, exploiting the symmetries of the diagrams simplifies the results considerably. We have provided a \texttt{Mathematica} file with the four and five point results; we hope that our results may serve as data points from which further insight can be obtained.

In the last part of our paper we discussed the flat space limit of our results. It is a nice feature of the momentum space formalism that this limit can be taken rather easily, and it is natural to wonder if such a transparent limit can help us find analogous structures between graviton and gluon to those found in flat space scattering amplitudes. For instance, KLT-like relations have been carried out in the context of cosmology \cite{Li:2018wkt}. Similarly, double copy like relations have been explored and are successfully realized for three point CFT correlators \cite{Farrow:2018yni}. It is an interesting question whether these analyses  can be further developed for holographic settings, and we hope that our work can assist in this direction.

Lastly, we hope our investigation can connect with \emph{cosmological bootstrap program} as AdS and dS correlators are closely related. For instance, in \cite{Sleight:2019mgd, Sleight:2019hfp} authors propose a framework for the computation of late-time correlators in de Sitter space and bridge the gap with the boundary correlators in anti-de Sitter space. Moreover, it has been shown that the wave function of the universe for conformally coupled scalar theories in $dS_4$ can be expressed as volumes of polytopes in the same spirit as the amplituhedron \cite{Benincasa:2018ssx, Arkani-Hamed:2018bjr, Arkani-Hamed:2017fdk}. Since, the structures that we have calculated have a similar flavor, it is possible that they have a polytopic interpretation. It would be interesting to explore these connections. 

\section*{Acknowledgments}
We thank Chandramouli Chowdhury for discussions. SA is supported by NSF grant PHY-1350180 and Simons Foundation grant 488651. SK was supported by DRFC Discretionary Funds from Williams College.

\appendix
\section{Scalar factor for five point diagram}
\label{appendix}
The scalar factor of the amplitude takes following form
\begin{equation*}
\mathcal{M}=\frac{\sum\limits_{i=0}^{9}k_3^i\left(c_i+(k_{\underline{12}}-k_{\underline{45}})d_i\right)}{2k_{12\underline{12}}^2k_{45\underline{45}}^2k_{\underline{12}3\underline{45}}^2k_{123\underline{45}}^3k_{\underline{12}345}^3k_{12345}^4}\;,
\end{equation*}
where $c_i$ ($d_i$) carries the even (odd) part of the amplitude under the exchange \mbox{$\{k_1,k_2\}\leftrightarrow\{k_4,k_5\}$}. We can succinctly express these factors in parameters manifest under this symmetry:
\begin{align*}
e\equiv&k_1+k_2+k_4+k_5\;,\\
\alpha\equiv&(k_1+k_2)(k_4+k_5)\;,\\
\beta\equiv&k_1k_2+k_4k_5\;,\\
\gamma\equiv&k_1k_2k_4k_5\;,\\
\sigma\equiv&\abs{k_1+k_2-k_4-k_5}\;,\\
\lambda\equiv&\abs{k_1k_2-k_4k_5}\;,\\
\rho\equiv&k_{\underline{12}}+k_{\underline{45}}\;,\\
\delta\equiv&k_{\underline{12}}k_{\underline{45}}\;,\\
\kappa\equiv&\abs{k_{\underline{12}}-k_{\underline{45}}}\;.
\end{align*}
In these parameters $c_i$ and $d_i$ read as follows:

\scriptsize
\begin{align*}
	c_0=&\frac{1}{2} e \rho  (2 (\alpha +\delta )+e \rho ) \bigg(4 \gamma  \rho  \left((\alpha
	-\delta )^2+\alpha  \rho ^2\right)+2 \delta  e^5 (\beta +\delta )+e^4 (2 \delta  (\rho 
	(2 (\alpha +\beta )+\delta )-\lambda  \sigma )+\alpha  \beta  \rho )\\&+e^3 \left(2 \delta 
	\left(\alpha ^2-2 \alpha  (\beta +\delta )+2 \beta  \delta +4 \gamma +\delta ^2\right)+2
	\rho ^2 (\delta  (2 \alpha +\beta )+\alpha  (\alpha +\beta ))-\alpha  \lambda  \rho 
	\sigma -4 \delta  \lambda  \rho  \sigma \right)\\&+e^2 \rho  \left(2 \alpha ^3-\alpha ^2
	\left(\beta +4 \delta -2 \rho ^2\right)+\alpha  \left(2 \delta  (\delta -\beta )+\beta 
	\rho ^2+4 \gamma -2 \lambda  \rho  \sigma \right)+\delta  (3 \beta  \delta +16 \gamma -2
	\lambda  \rho  \sigma )\right)\\&-e \rho  \left(\alpha ^2 \lambda  \sigma +\alpha 
	\left(\lambda  \sigma  \left(\rho ^2-2 \delta \right)-8 \gamma  \rho \right)+\delta 
	(\delta  \lambda  \sigma -8 \gamma  \rho )\right)\bigg)
\\
	c_1=&\left(4 \rho  \left(\alpha +\delta +e^2\right)+2 e (\alpha +\delta )+5 e \rho ^2\right)
	\bigg(4 \gamma  \rho  \left((\alpha -\delta )^2+\alpha  \rho ^2\right)+2 \delta  e^5
	(\beta +\delta )+e^4 (2 \delta  (\rho  (2 (\alpha +\beta )+\delta )-\lambda  \sigma
	)+\alpha  \beta  \rho )\\&+e^3 \left(2 \delta  \left(\alpha ^2-2 \alpha  (\beta +\delta )+2
	\beta  \delta +4 \gamma +\delta ^2\right)+2 \rho ^2 (\delta  (2 \alpha +\beta )+\alpha 
	(\alpha +\beta ))-\alpha  \lambda  \rho  \sigma -4 \delta  \lambda  \rho  \sigma
	\right)+e^2 \rho  \Big(2 \alpha ^3-\alpha ^2 \left(\beta +4 \delta -2 \rho
	^2\right)\\&+\alpha  \left(2 \delta  (\delta -\beta )+\beta  \rho ^2+4 \gamma -2 \lambda 
	\rho  \sigma \right)+\delta  (3 \beta  \delta +16 \gamma -2 \lambda  \rho  \sigma
	)\Big)-e \rho  \left(\alpha ^2 \lambda  \sigma +\alpha  \left(\lambda  \sigma 
	\left(\rho ^2-2 \delta \right)-8 \gamma  \rho \right)+\delta  (\delta  \lambda  \sigma -8
	\gamma  \rho )\right)\bigg)
\end{align*}
\normalsize

\includepdf[pages=-,pagecommand={},width=\paperwidth]{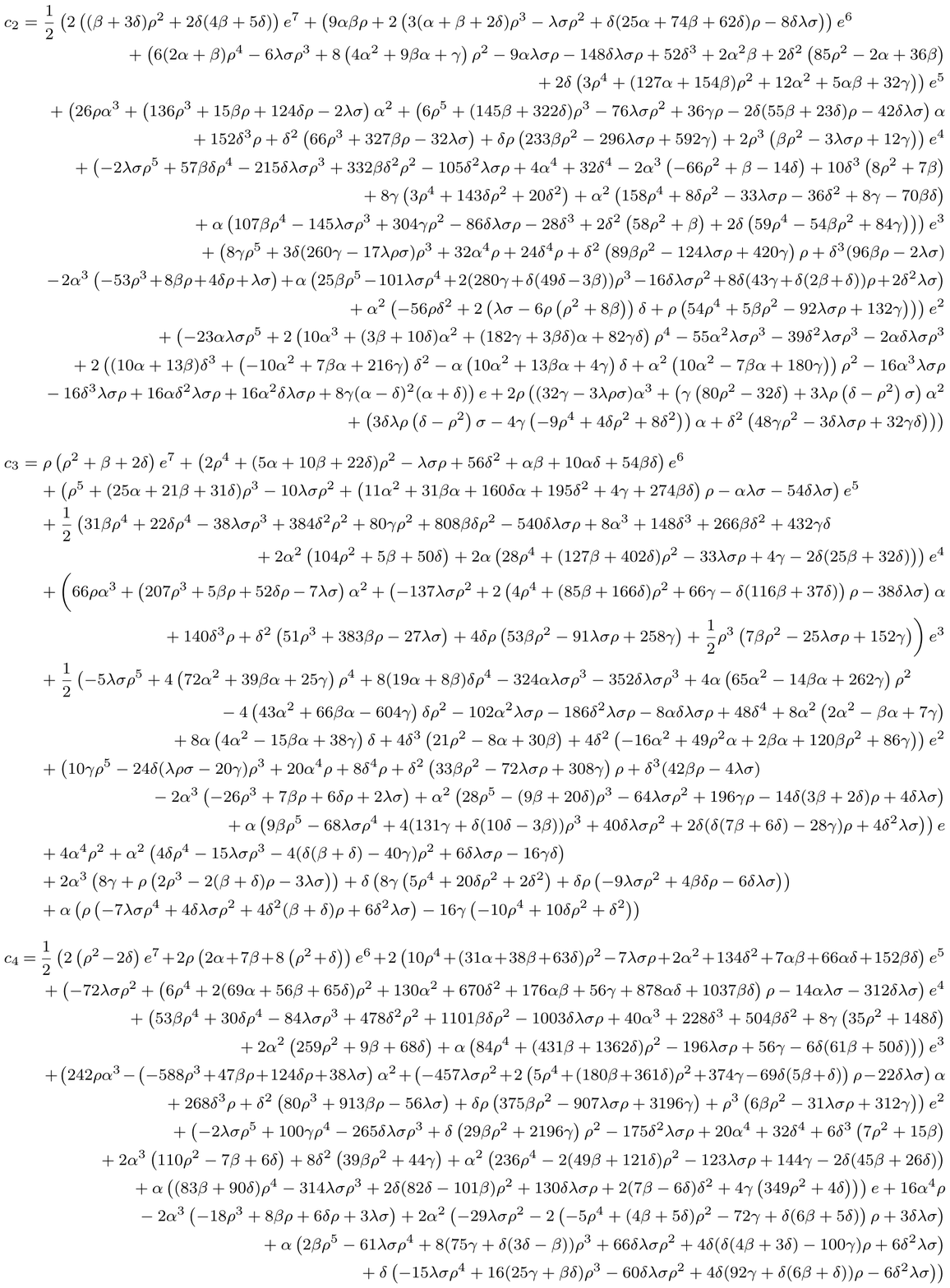}
\bibliographystyle{utphys}
\bibliography{references}{}
\end{document}